\documentclass[prc,showpacs,preprintnumbers,amsmath,amssymb]{revtex4}
\usepackage{graphicx}
\usepackage{graphicx}
\usepackage{dcolumn}
\usepackage{bm}
\begin{document}
\title
{
Effect of compactification of twisted toroidal extra-dimension 
on sterile neutrino}
\author{A. K. Mohanty}
\email{ajitkumar.mohanty@saha.ac.in}

\affiliation{Saha Institute of Nuclear Physics, Kolkata - 700064 and Bhabha Atomic Research Centre, Mumbai-400084, INDIA}
\date{\today}

\begin{abstract}
{
We consider a toroidal extra-dimensional space with shape moduli $\theta$ which is the angle between
the two large extra dimensions $R_1$ and $R_2$ (twisted LED with $\delta=2$). 
The Kaluza-Klein (KK) compactification 
results in a tower of KK bulk neutrinos which are sterile in nature and couple to
the active neutrinos in the brane. The active-sterile mixing probability strongly depends on the angle $\theta$
due to changing pattern of KK mass gaps which leads to level crossing.
Considering only the first two lowest KK states in analogy with $(3+2)$ model, it is shown that 
$|U_{\alpha 4}| > 
|U_{\alpha 5}|$ when $\theta = \pi/2$ corresponding to the case of a normal torus.
Since $\Delta_{14}^2 < \Delta_{15}^2$, this is expected in normal LED model as higher the sterile mass lower is the mixing probability.
Contrary to this expectation, it is found that
there exists a range in $\theta$ where $|U_{\alpha 5}| \ge |U_{\alpha 4}|$ even though $\Delta m_{14}^2 <
\Delta m_{15}^2$ which has been demonstrated qantitatively using fourier transformation of reactor anti-neutrino spectrum.
This is an important observation 
which can be linked to the oscillation parameters extracted by several
$(3+2)$ global analyses of the neutrino and anti-neutrino data obtained from the short base line  
measurements.      
}
\end{abstract}

\maketitle

\section {Introduction}

We consider the compactification of toroidal two extra dimensions characterised by a shape moduli $\theta$
which is the angle bewteen the two large extra dimensions $R_1$ and $R_2$. The non trivial effect of shape
moduli on compactification was first studied by Dienes and Mafi who brought out
a number of profound phenomena relevant for the interpretation of experimental data if such extra dimensions
exist \cite{Dienes_prl1,Dienes_prl2}. Notably among them is the changing pattern of Kaluza-Klein (KK) mass gaps which strongly depend on
$\theta$ and exhibit level crossing making some of the higher KK modes lighter as compared to the lower ones
when $\theta$ is varied. This is an important aspect which we incorporate in the  
ADD (Arkani-Hamed, Dimopoulos and Dvali) model with two large extra dimensions (LED) to study the active-sterile neutrino
mixing  in $4+\delta$ dimensions with  $\delta=2$
\cite {ADD1,ADD2}. Another impetus to this work stems from the recent observation that
the fit to the short base line (SBL) reactor anti-neutrino measurements with new anti-neutrino flux \cite{muller, huber} 
improves considerably if two sterile neutrions $(3+2)$ are assumed instead of one $(3+1)$ \cite{Kopp_prl}. Naively, it is
observed that $\Delta m_{51}$ is about $\sqrt{2}$ times larger than $\Delta m_{41}$ and $|U_{15}| \ge |U_{14}|$ \cite{Kopp_prl,Kopp1}.
The LED model with $\delta=1$ (one extra dimension larger than others) results in a tower of KK sterile neutrions with
KK mass increasing as $n$ and mixing probability decreasing as $1/n^2$ where $n=1,2,3$ etc
\cite{Anton, ADD3, Dienes, Dvali, Bar, Davo, Gin, Cao, Gonzalez, Rode}. Obviously, the case with $\delta=1$ is not
consistent with the above observations if we consider first two lowest sterile states. In case of $\delta=2$, the KK mass 
increases as $\sqrt{m^2+n^2}$ where $m$ and $n$ are two different KK modes associated with $R_1$ and $R_2$ respectively.
Although the mass of first two KK modes $(1,0)$ or $(0,1)$  and $(1,1)$  differ by a factor of $\sqrt{2}$, 
as expected, still it does not predict the observed active-sterile mixing
probability when $\theta=\pi/2$. On the other hand, when $\theta$ is close to $\pi/4$, the predicted masses and mixing probabilities
are found to be consistent with the above experimental obsevations. In this letter,  we consider compactification
on a general two-torus corresponding to $\delta=2$ instead of $\delta=1$ as
one dimensional compactification lacks shape moduli. It is shown here
that there exists a range in $\theta$ where $|U_{\alpha 5}| \ge |U_{\alpha 4}|$ even though $\Delta m_{14}^2 <
\Delta m_{15}^2$. This is an important observation  which is demonstrated more quantitatively using cosine
Fourier transformation  of the reactor anti-neutrino spectra.

\section{Formalism}

We consider a brane world theory with $6$ dimensional bulk, 
where the active neutrinos are confined to the brane and the singlet sterile neutrino
$\Phi^\alpha(x^\mu,y_1,y_2)$ 
propagates in the bulk with extra dimensions $y_1$ and $y_2$. 
Using the Kaluza-Klein (KK) expansion, the singlet fermionic field can be
expanded as,

\begin{eqnarray}
\Phi_{R/L}(x^\mu,y)\sum_m \sum_n \Phi_{R/L}^{(m,n)}(x^\mu)~f^{mn}(y_1,y_2),
\label{p1}
\end{eqnarray}
where $\mu=0,1,2,3$ are co-ordinates belonging to the brane and $y_1,~y_2$ are the co-ordinates of two extra dimensions. The
subscripts $R$ and $L$ refer explicitly to four dimensional Lorentz property.
The periodic function $f^{mn}(y_1,y_2)$ is given by,

\begin{eqnarray}
f^{mn}(y_1,y_2)=\frac{1}{\sqrt{V}}exp \left [ i \frac{m}{R_1} \left ( y_1-\frac{y_2}{tan \theta} \right ) 
+  i \frac{n}{R_2} \frac{y_2}{sin \theta} \right ],
\label{f1}
\end{eqnarray}

with periodicity $y_1 \sim y_1+ 2 \pi (R_1+R_2 cos \theta)$ and $y_2 \sim y_2+2 \pi R_2 sin \theta$ \cite{Dienes_prl1}.
The normalization factor $V=4 \pi^2 R_2 sin \theta (R_1+R_2 cos \theta)$ plays the role of volume of the extra-dimensions.
Note that for $\theta=\pi/2$, $V=4 \pi^2 R_1 R_2$ which is the volume of a normal torus. Eq. \ref{f1} satisfies the condition,

\begin{eqnarray}
\frac{1}{V}\int_0^{\infty} \left ( f^{pq} \right )^{*}  f^{mn} dy_1 dy_2 =\delta_{pm}\delta_{qn}.
\label{f2}
\end{eqnarray}

The bulk action responsible for the neutrino mass is given by (kinetic term is not included) \cite{Dudas},  

\begin{eqnarray}
A_{bulk}= \int d^4x~dy_1~dy_2~
\big[
\Phi_L^\dagger (\partial_5+ i \partial_6) \Phi_R 
-\Phi_R^\dagger (\partial_5-i \partial_6) \Phi_L
\big ].
\label{s1}
\end{eqnarray}
Using Eq. \ref{p1}, Eq. \ref{f1}, Eq. \ref{f2}  and the substitution,

\begin{eqnarray}
\Psi_R^{0,0} = \Phi_R^{0,0}~~;~~\Psi_R^{m,n}=\frac{1}{\sqrt{2}} \left ( \Phi_R^{m,n}+\Phi_R^{-m,-n} \right )~~;~~
\Psi_L^{m,n}=\frac{1}{\sqrt{2}} \left ( \Phi_L^{m,n}+\Phi_L^{-m,-n} \right ),
\label{p2}
\end{eqnarray}
the $y_1$ and $y_2$ variable in Eq. \ref{s1} can be integrated out to get,

\begin{eqnarray}
A_{bulk}=-\int d^4x
\sum_{m,n}^N \frac{k_{m,n}}{ R} \big ( \Psi_R^{(m,n)\dagger} \Psi_L^{(m,n)}+\Psi_L^{(m,n)\dagger} \Psi_R^{(m,n)} \big ) ,
\label{s2}
\end{eqnarray}
where the absolute value of the mass term for $(m,n)$ mode is given by,

\begin{eqnarray}
k_{m,n} =\sqrt{\frac{1}{sin \theta} \left (m^2 + n^2 -2 m~n~
 cos \theta \right )}.
\label{k1}
\end{eqnarray}

We have relaxed the condition further by assuming that $R_1=R_2=R$.
Note that the summation $\sum_{mn}$ above is over all
modes of $m$ and $n$ upto a maximum value of $N$, but excluding $m=n=0$ mode.
We can now add  the relevant portion of interaction term between brane and the bulk field containg mass,

\begin{eqnarray}
A_{int}=-m_D \int d^4 x  \bigg [ \nu_L^\dagger   
       \bigg ( \nu_R+\sqrt {2} \sum_{m,n}^N \Psi_R^{(m,n)}    \bigg ) +hc \bigg ]
\label{s3}
\end{eqnarray}
where $\nu_R=\Psi_R^{(0,0)}$ and $m_D$ is the Dirac neutrino mass generated due to coupling of bulk neutrinos with the brane localized
SM Higgs boson at $y_1=y_2=0$. 
Finally, by collecting the neutrino mass terms in the Lagrangian and explicitly including the neutrino flavor indices $\alpha$ and
$\beta$, we obtain,

\begin{eqnarray}
\mathcal{L}_{mass}=- \sum_{\alpha=1}^3 \sum_{m, n}^N \frac{k_{m,n}}{R} \Psi_R^{\alpha (m,n) \dagger} \Psi_L^{\alpha (m,n)}      
-\sum_{\alpha, \beta=1}^3 m_D^{\alpha \beta} \bigg (\nu_R^{\alpha \dagger} +
\sqrt 2 \sum_{m,n}^N \Psi_R^{\alpha (m,n) \dagger} \bigg) \nu_L^\beta + hc
\label{l}
\end{eqnarray}

Note that while the right handed sterile neutrino $\Psi_R^{0,0}$ participates in the process of mass generation at the
brane, the left handed sterile neutrino $\Psi_L^{0,0}$ decouples from the mass part of the Lagrangian as $k_{mn}$ vanishes
for $(0,0)$ mode. We also neglect the Majrona mass and associate suitable lepton number to $\Psi_R$ so that
lepton number is conserved. 
The formalism is now similar to the case of $\delta=1$ and can be found in several works 
\cite{Anton, ADD3, Dienes, Dvali, Bar, Davo, Gin, Cao, Gonzalez, Rode}. Therefore, following the standard procedure of
diagonalizing the Dirac mass term $m^{\alpha \beta}$ with PMNS matrx $U$ and making a symmetric transformation to a set of new
basis, the mass Lagrangian can be written in a compact form given by \cite{Dienes, Cao},

\begin{eqnarray}
\mathcal{L}_{mass}=\frac{1}{2}\bigg (\nu^\dagger \it{M} \nu + hc \bigg ),
\label{L6}
\end{eqnarray}

where the neutrino mass matrix $M$ is given by,

\begin{eqnarray}
M=\left (\begin{array} {ccccccc}
0  & m_\nu & m_\nu  & m_\nu & m_\nu & \cdots & m_\nu \\
m_\nu  &0 & 0  & 0 &0 & \cdots &0 \\
\vdots & \vdots & \vdots & \vdots& \vdots &\vdots& \vdots\\
m_\nu  & 0 & \frac{k_{mn}}{R} & 0 & 0 &\cdots &0 \\
m_\nu  & 0 & 0 & -\frac{k_{mn}}{R} & 0 & \cdots &0 \\

\vdots & \vdots & \vdots & \vdots& \vdots &\vdots& \vdots\\
m_\nu & 0 & 0 & 0 &\cdots &\frac{k_{NN}}{R} &0 \\
m_\nu & 0 & 0 & 0 &0 &\cdots &  -\frac{k_{NN}}{R}\\
\end{array} 
\right ), 
\label{M3}
\end{eqnarray}
with $\nu=(\nu_L, \nu_R, \nu_s^{-1}, \nu_s^{1} \cdots)^T$. In the above, the
Dirac mass $m_\nu$ could be either $m_1$, $m_2$ or $m_3$ depending on whether it is $e$, $\mu$ or $\tau$ neutrino.
The mass term $k_{mn}/R$ appearing in Eq. \ref{M3}
represents  ($d_k \times d_k$) block diagonal matrix, $d_k$ being the degeneracy of the $(m,n)$
mass state, $N$ is the upper limit of $m$ or $n$.  
For $N=1$, the independent modes are $(1,0)$ and $(1,1)$ since $(0,1)$ and $(1,0)$ mode degenerates.
Similarly for $N=2$, the independent modes are  $(0,1)_{d=2}$, $(1,1)$, $(2,0)_{d=2}$,
$(1,2)_{d=2}$ and $(2,2)$. Therefore, for the case of $\delta=2$, $d$ is either $2$ or $1$. Finally, we will
make a distinction between $k_{mn}$ and $k$ which will be used interchangably in the following.
The $k_{mn}$ represents a mass state as defined in Eq. \ref{k1} for a given
$(m,n)$ mode where as the index $k$ represents the $k^{th}$ state. In the above example of $N=2$, the six states
including a $(0,0)$ mode can be represented by the index $k=0,1,2,3,4,5$ and each state having mass given by
$k_{mn}$ with degeneracy $d_k$.

\section{Eigen value and Eigen Vector}

The eigen value $\lambda$ of Eq. \ref{M3} can be obtained from the
characteristic equation, $Det(M-\lambda I)=0$ given by \cite{Cao},

\begin{eqnarray}
\bigg [ \prod_{m,n}^N \bigg (\frac{k_{mn}^2}{R^2}-\lambda^2 \bigg )^{d_{k}} \bigg ]
\bigg [ \lambda^2-m_\nu^2+2\lambda^2 m_\nu^2 \sum_{m,n}^N \frac{d_{k}}{\frac{k_{mn}^2}{R^2}-\lambda^2} \bigg ] = 0. 
\label{E1}
\end{eqnarray}

There are ($d_{k}-1$) states for which $\lambda$ is equal to $k_{mn}/R$ and one state for which
$k_{mn}$ is not equal to $k_{mn}/R$ for which the solutions can be obtained from,

\begin{eqnarray}
\bigg [ \lambda^2-m_\nu^2+2\lambda^2 m_\nu^2 \sum_{m,n}^N \frac{1}{\frac{k_{mn}^2}{R^2}-\lambda^2} \bigg ] = 0. 
\label{E2}
\end{eqnarray}

In Eq. \ref{E2}, the factor $d_{k}$ is not included explicitly as the summation over $(m,n)$ takes care of it.
Unlike $\delta=1$ case, the summation in the above equation is logarithmically divergent.
Therefore, we solve for a given $\lambda_k$ iteratively up to a cut-off scale set by $k_{NN}/R$. 

The matrix Eq. \ref{M3}  can also be diagonalised by the unitary matrix $L$ whose $k^{th}$ column matrix corresponding to
the mode $(m,n)$ is given by,

\begin{eqnarray}
L^{k}
=\frac{1}{\sqrt{B}} \bigg (1, \frac{m_\nu}{\lambda_k},
\cdots, \frac{m_\nu}{\lambda_k-\frac{1}{R}}, \frac{m_\nu}{\lambda_k+\frac{1}{R}}, \cdots,  \frac{m_\nu}{\lambda_k-\frac{k_{mn}}{R}},
\frac{m_\nu}{\lambda_k+\frac{k_{mn}}{R}}, \cdots, 
\frac{m_\nu}{\lambda_{k}-\frac{k_{NN}}{R}},\frac{m_\nu}{\lambda_{k}+\frac{k_{NN}}{R}} \big )^T 
\label{U1}
\end{eqnarray}
The normalization factor $B$ is obtained from the condition $(L^k)^T L^k=1$. 

The neutrino state of a given flavor $\nu^\alpha_L$  can be written in terms of mass eigen states as,

\begin{eqnarray}
\nu_L^\alpha = \sum_{j=1}^{3} U^{\alpha j} \sum_{k} L_j^{0k} \nu_L^{\prime j (k)},
\label{nu1}
\end{eqnarray}
where $(L_j^{0k})^2=2/B$.
A quantity of crucial interest is the survival probability of neutrino of flavor $\alpha$ after
travelling a distance of $L$ is given by \cite{Davo},
 
\begin{eqnarray}
P_{\alpha \alpha}(L)=\left |\sum_{j=1}^{3} \left |U^{\alpha j} \right |^2 
\sum_{k} (L_j^{0k})^2 exp \left (i\frac{2.54 \lambda_j^{(k)2} L}{ E_\nu } \right ) \right |^2, 
\label{p3}
\end{eqnarray}

where $E_\nu$ is the neutrino energy in MeV, the eigen value $\lambda_j$ is in eV and $L$ is in $m$. In Eq. \ref{p3}, the
subscript $j=1,2,3$ refers to $e, \mu, \tau$ respectively.

In analogy with $(3+n)$ model, we can define another parameter of interest $S_{\alpha k}$ as,

\begin{eqnarray}
S_{\alpha k}= \sum_{j=1}^3 \left |U_{\alpha j} L_j^{0k} \right |^2,
\label{s4}
\end{eqnarray}

so that we can
identify the parameters $S_{\alpha 1}$  and $S_{\alpha 2}$ with either $ |U_{\alpha 4}|^2$ or $|U_{\alpha 5}|^2$.
Note that in the absence of extra dimensions,
$S_{\alpha 0}$ is equal to unity since PMNS matrix is unitary. However, it is less than unity when $L_j^{0k}$
is included. For neutrino mass square differences, we use 
$\Delta m_{21}^2=7.45 \times 10^{-5}$ $eV^2$, $\Delta m_{31}^2=2.417 \times 10^{-3}$ $eV^2$ \cite{Garcia}. In the normal hierarchy
(NH: $m_3 > m_2 >m_1$), we consider $m_1$ as a variable and
express  $m_2{^2}=m_1^2+\Delta m_{21}^2$ and $m_3^2=m_2^2+\Delta m_{31}^2$ respectively. In the
inverted hierarchy (IH: $m_2 > m_1 >m_3$ ), $m_3$ is considered as a variable and express $m_2^2=m_3^2+\Delta m_{32}^2$ and 
$m_1^2=m_3^2+\Delta m_{31}^2$ respectively. Other parameters are
$sin^2 \theta_{12}=0.313$, $sin^2 \theta_{23}=0.444$ and $sin^2 \theta_{13}=0.0244$ respectively \cite{Garcia}.

\begin{figure}
\begin{center}
\includegraphics[scale=.6]{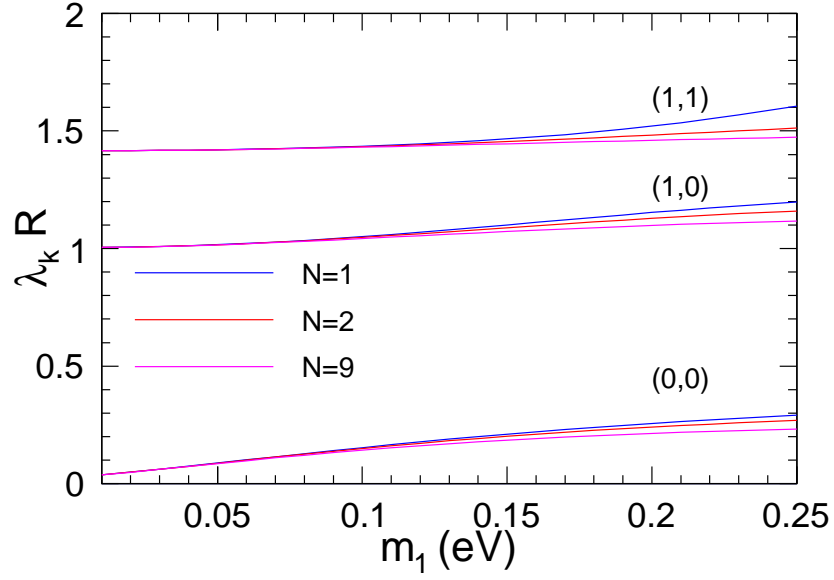}
\caption{The figure shows the plots of $\lambda_k R$ as a function of $m_1$ for $\theta=\pi/2$ and 
for a fixed value of $R=3.1 \times 10^{-7}$ m.
Here, $k$ refers to the $k^{th}$ eigen value corresponding to a given $(m,n)$ mode. The eigen value
$\lambda_k$ is defined as $\frac{1}{3}\sum_j \lambda_k^j$.}
\label{fig_2d_1}
\end{center}
\end{figure}

The fig. \ref{fig_2d_1} shows the plot of eigen values $\lambda_k R$ as a function of mass $m_1$ for $N=1,2,9$
(other masses are fixed
based on NH) solved iteratively using the Eq. \ref{E2}. The eigen value $\lambda_k$ is defined as 
the average $\frac{1}{3}\sum_j \lambda_k^j$.  
It is easy to check that Eq. \ref{E2} has large
number of solutions $\lambda_{k}$ depending on the $m$ , $n$ and  $N$ values, although shown for
only $(0,0)$, $(1,0)$ and $(1,1)$ modes only since $(0,1)$ and $(1,0)$ modes are degenerate.
As can be seen, the solutions are
not convergent due to logarithmic divergent and  strongly depend on the choice of cut-off value $N$ 
particularly for large $m_1$ values. However,
for small $m_1$ (more specifically when $\xi_j=m_\nu^j R << 1$), the dependency on $N$ is rather weak and the solutions
for $m$ or $n$ $ \ge 1$, can be approximated by \cite{Dienes,Cao},

\begin{eqnarray}
\lambda_k=\frac{k_{mn}}{R} \big (1+\frac{\xi^2_j}{k_{mn}^2}-\frac{\xi^4_j}{k_{mn}^4}+.... \big ).
\label{L7}
\end{eqnarray}

\begin{figure}
\begin{center}
\includegraphics[scale=.6]{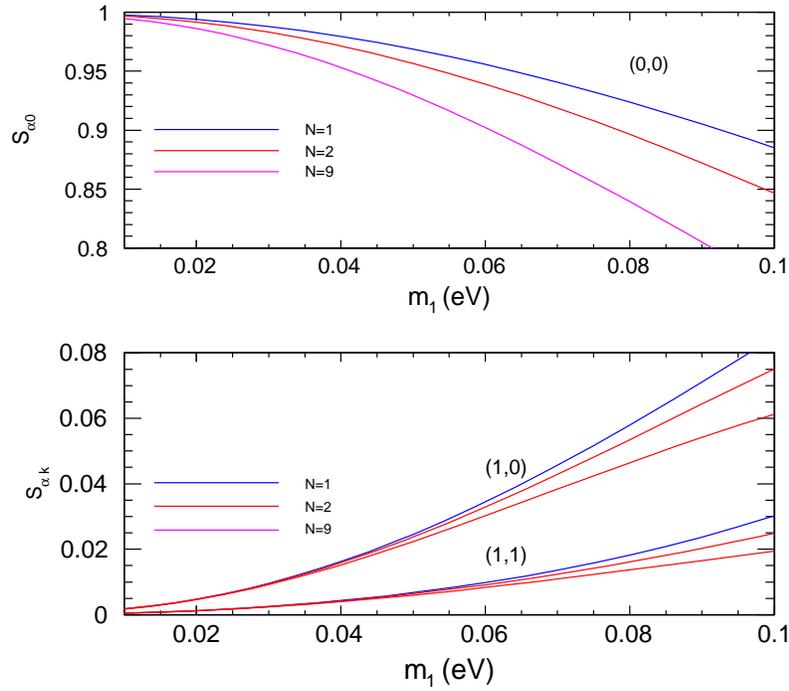}
\caption{The top panel shows the plots of 
$S_{\alpha 0}$ for $(0,0)$ mode as the  function of $m_1$ for $N=1$, $N=2$ and $N=9$ respectively.
The bottom panel shows the similar plots for $(1,0)$ and $(1,1)$ modes. The parameters used are same as 
that of fig. \ref{fig_2d_1}. }

\label{fig_2d_2}
\end{center}
\end{figure}

The top panel of fig. \ref{fig_2d_2} shows the plots of $S^{\alpha 0}$ as given in Eq. \ref{s4}
as the function of $m_1$ for $N=1$, $N=2$ and $N=9$ respectively for $(0,0)$  mode.
It is noticed that $S_{\alpha 0}$ is close to unity when $m_1$ is very small as expected, however unitarity is
violated with increasing $m_1$. The unitarity is violated by more than $5\%$ at $m_1=0.05$ eV even when
$N=2$. 
The bottom pannel shows the similar plots for $(1,0)$ and $(1,1)$ modes respectively. As can be seen,
the value of $S_{\alpha k}$ strongly depends on $N$ at large values of $m_1$ even though  $S_{\alpha 0}$ starts decreasing
significantly with increasing $N$. Since $S_{\alpha k}$ is not very sensitive to $N$ for small mass,
the mixing probability $(L_j^{0 k})^2$ (hence $S_{\alpha k}$)  can be approximated by,

\begin{eqnarray}
(L_j^{0k})^2=\frac{2 d_k \xi_j^2}{k_{mn}^2}.
\label{l1}
\end{eqnarray}

\begin{figure}
\begin{center}
\includegraphics[scale=.6]{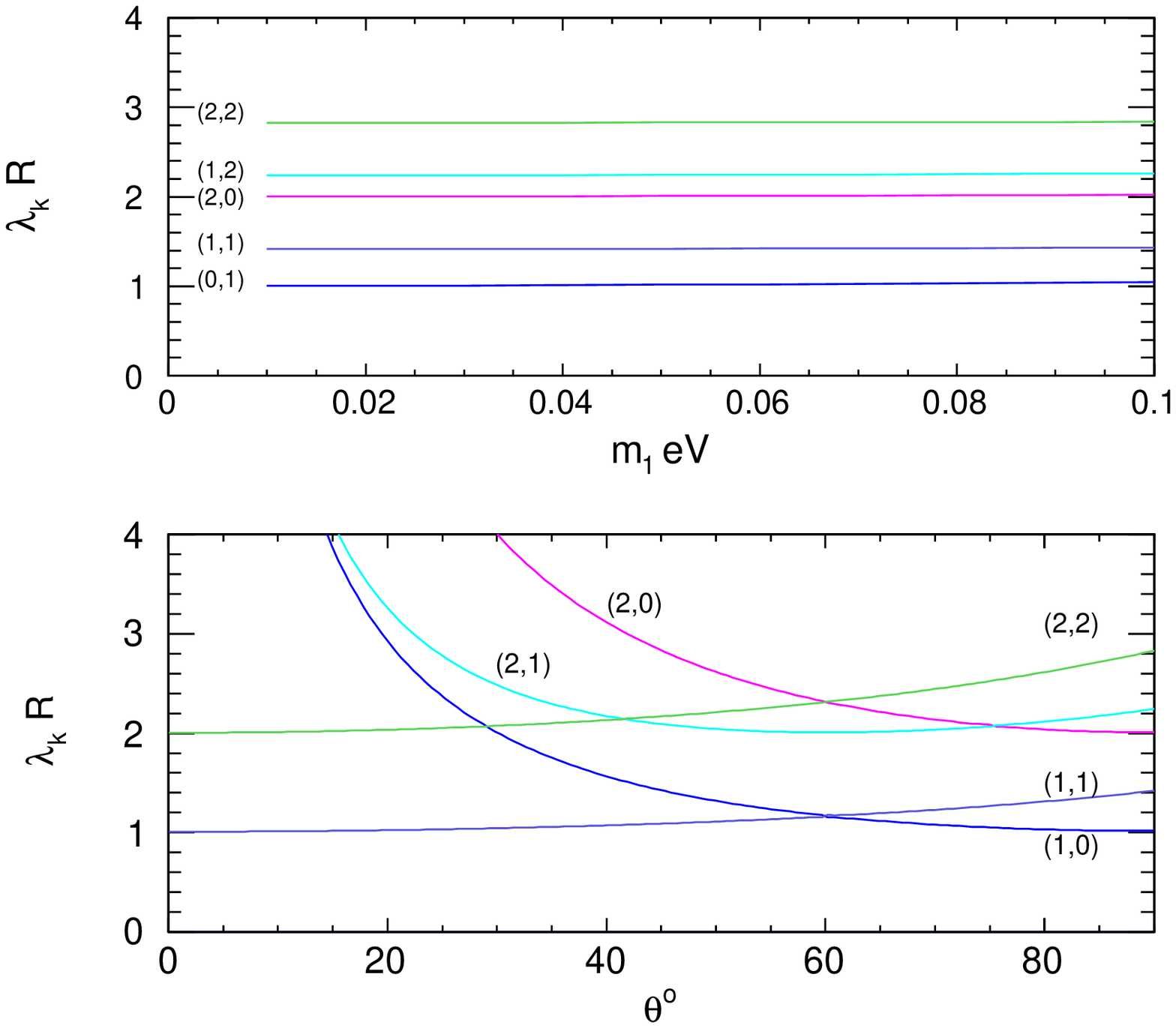}
\caption{The top panel shows the plots of 
$\lambda_k R$ as a function of $m_1$ for $(1,0)$, $(1,1)$, $(2,0)$, $(2,1)$ and $(2,2)$ modes 
corresponding to $N=2$, $\theta=\pi/2$ and $R=3.1 \times 10^{-7}$. The bottom panel shows the similar
plots as a function of $\theta$ at a fixed mass $m_1=0.052$ eV.}

\label{fig_2d_extra}
\end{center}
\end{figure}

\begin{figure}
\begin{center}
\includegraphics[scale=.6]{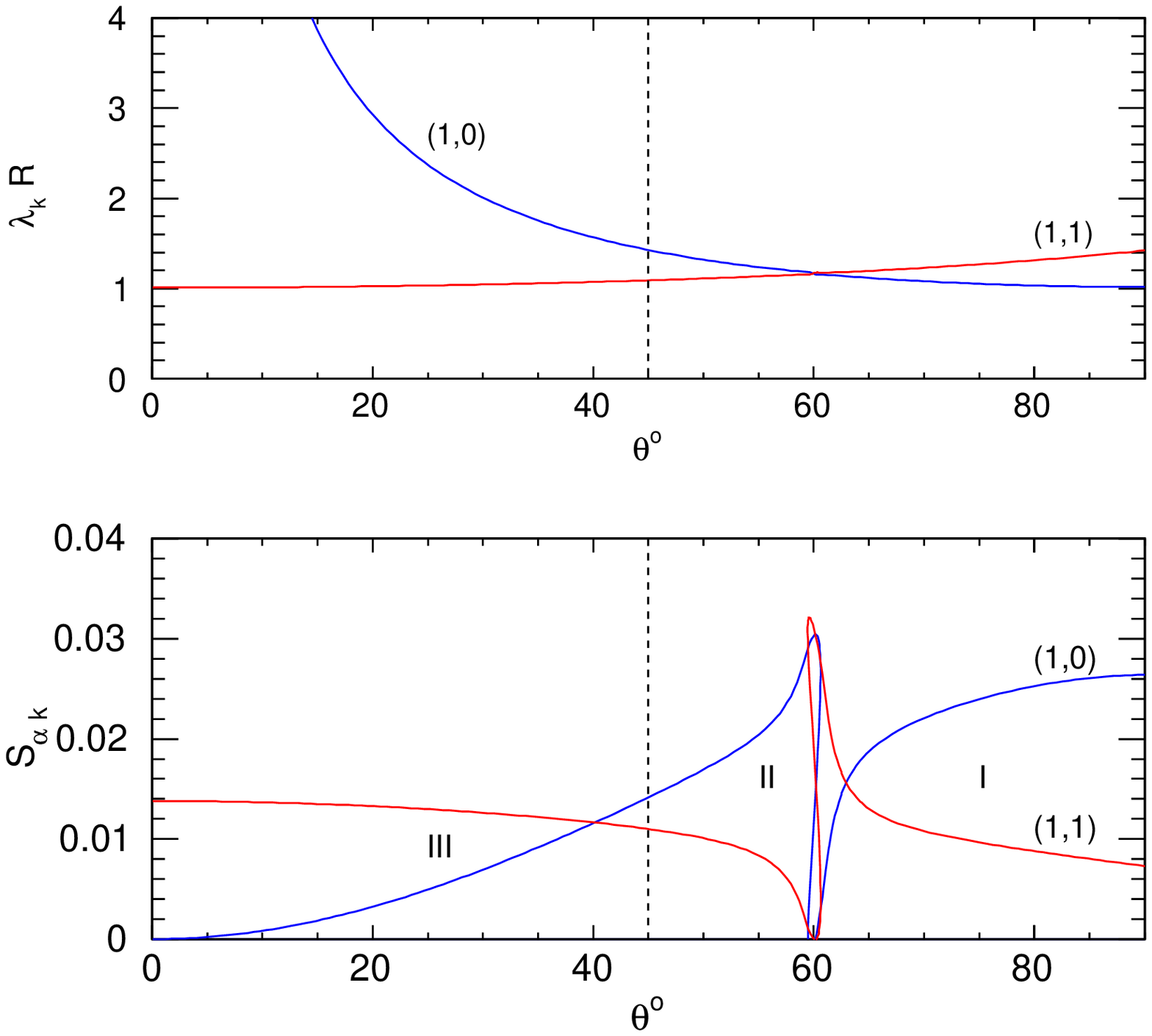}
\caption{The top panel shows the plot of $\lambda_k R$ as a function of $\theta$ for $(1,0)$ and $(1,1)$ modes.
The other parameters are $N=1$, $m_1=0.052$ eV and $R= 3.1 \times 10^{-7}$ m. The bottom panel shows the corrresponding
$S_{\alpha k}$ as a function of $\theta$.}  

\label{fig_2d_3}
\end{center}
\end{figure}

The top panel of fig. \ref{fig_2d_extra} shows 
the plot of $\lambda_k R$ as a function of $m_1$ for a few lowest mass states
corresponding to $N=2$. This corresponds to the case of normal torus for which $\theta=\pi/2$. The other
parameter values are listed in the figure caption. For small values of $m_1$, $\lambda_k R$ is nearly equal to 
$k_{mn}$ as expected. The ratio of the mass gaps with respect to the lowest one are $\sqrt{2}$, $2$, $\sqrt{5}$,
$2\sqrt{2}$ respectively. 
The bottom panel shows the similar plot but as a function of $\theta$ at a fixed mass $m_1=0.52$
eV. The pattern of KK mass gaps now change with decreasing $\theta$ and exhibit level crossing making some
of the higher modes lighter as compared to lower ones. Although this phenomena has been studied in 
detail before \cite{Dienes_prl1,Dienes_prl2}, we consider here only the first two mass states $(1,0)$ and $(1,1)$
which shows level crossing for $\theta < \pi/3$ which is shown more specifically in fig. \ref{fig_2d_3} (see top panel).
The bottom panel shows the active-sterile mixing probabilities $S_{\alpha 1}$ and $S_{\alpha 2}$ as a function of 
$\theta$. In the region I, mass of $(1,1)$ mode is higer than $(1,0)$ mode and in the region III, the mass of $(1,0)$ mode
is higher than $(1,1)$ mode. Accordinly, the mixing probability $S_{\alpha 1} > S_{\alpha 2}$ in region I and
$S_{\alpha 2} > S_{\alpha 1}$ in region III as expected. However, the behavior is different in region II where 
$S_{\alpha 1} > S_{\alpha 2}$ even though the mass of $(1,0)$ mode is heavier than the mass of $(1,1)$ mode which is contrary to the naive
expectation. 
For $\theta < 60^o$, although the mass of $(0,1)$ mode
becomes higher than the mass of $(1,1)$ mode, the $(0,1)$ mode has degeneracy two times higher than $(1,1)$ mode.
So the net result is $S_{\alpha 1}$ remains higher than $S_{\alpha 2}$ for small values of $\xi$ (see Eq. \ref{l1}). 
It can be seen that at around $\theta \sim 40^o$, the two mixing probabilities are nearly equal
as the mass of $(1,0)$ mode 
is nearly $\sqrt{2}$ times
higher than mass of $(1,0)$ mode.  
In general, $S_{\alpha 1} \ge S_{\alpha 2}$ in the range $40^o < \theta < 60^o$.
Associating the mass of $(1,0)$ mode to $\Delta m_{15}$ and mass of $(1,1)$ mode to $\Delta m_{14}$ in the region II,
it would mean $|U_{e5}| \ge |U_{e4}|$. This is an interesting observation indicating that there exists a range
in $\theta$ where the mixing probability may become higher for havier mass and can be verified experimentally. In the
present study, we have three parameters $m_1$, $R$ and $\theta$. While $R$ decides the mass scale, $m_1$ controls the
mixing probability and the angle $\theta$ decides the relative strength of  the active-sterile coupling strength.
Although, we do not optimize the above parameters to explain experimental observations, we notice that the choice of
$R \sim 3.1 \times 10^{-7}$ m, $m_1=0.052$ $eV$ and $\theta=\pi/4$ describes the experimental observations reasonably well.

\begin{table}[ht]
\caption{The extracted parameters using $m_1=0.052$ eV and $R=0.31 \times 10^{-7}$
$m$ both for NH and IH.} 

\begin{center}
\begin{tabular}{llllll}
\hline
\hline
Type  &~~~Angle &~~$\Delta m_{14}^2~(eV^2)$ &~~$\Delta m_{15}^2~(eV^2)$  &~~$|U_{e4}|$ &~~$|U_{e5}|$ \\
\hline
 NH &~~~$90^o$ &~~0.42 &~~0.82 &~~0.160 &~~0.09 \\
\hline
 IH &~~~$90^o$ &~~0.42 &~~0.81 &~~0.220 &~~0.12 \\
\hline
 NH &~~~$45^o$ &~~0.48 &~~0.82 &~~0.105 &~~0.120 \\
\hline
 IH &~~~$45^o$ &~~0.48 &~~0.83 &~~0.137 &~~0.160 \\
\hline
\hline
\end{tabular}
\label{table1}
\end{center}
\end{table}

\begin{table}[ht]

\caption{The $(3+2)$ global fit parameters taken from [7].
The values in first row are extracted from reactor anti-neutrino data and the values in second row 
are exracted from global fits.}

\begin{center}
\begin{tabular}{llll}
\hline
\hline
$\Delta m_{14}^2~(eV^2)$ &~~$\Delta m_{15}^2~(eV^2)$  &~~$|U_{e4}|$ &~~$|U_{e5}|$ \\
\hline
0.46 & 0.89 & 0.108 & 0.124\\
\hline
0.47 & 0.87 & 0.128 & 0.138\\
\hline
\hline
\end{tabular}
\label{table2}
\end{center}
\end{table}

In table \ref{table1}, we have listed a few parameters estimated at $\theta=\pi/4$ and $\theta=\pi/2$
using both normal and inverted hierachy. The estimated values are compared with the reported results which are given in the table
\ref{table2}. The choice of $m_1=0.052$ $eV$ results in total active neutrino mass $\sum m_\nu=0.176$ eV 
which is less than the latest cosmological bound $\sum m_\nu<0.183$ $eV$  \cite{Giu}. Since inclusion of sterile neutrino will exceed
this upper bound, probably sterile neutrinos if present are not in thermal equilibrium 
in the cosmological context.

\section{Fourier transform of reactor anti-neutrino spectra}

The reactor anti-neutrino flux can be parametrized as the exponential
of a fifth order polynomial valid in the range $1.8 \le E \le 8$ MeV  \cite{huber,muller},

\begin{eqnarray}
\Phi(E)=exp \left ( \sum_{i=1}^{6} \alpha_i E^{i-1} \right ),
\end{eqnarray}

where $\alpha_i$s are listed in table \ref{table3}.

\begin{table}[ht]
\caption{The fit parameters for various isotopes that contribute to the total
power of the reactor. The parameters except for $U^{238}$ are taken from  
\cite{huber} and for $U^{238}$ from \cite{muller}.}
\begin{center}
\begin{tabular}{lllllllll}
\hline
\hline
& Isotope &$~~\alpha_0$ &~~$\alpha_1$ &~~$\alpha_2$ &~~$\alpha_3$ &~~$\alpha_4$ &~~$\alpha_5$&\\
\hline
&$U^{235}$ &~~4.367 &~~-4.577&~~2.100&~~-5.294(-1)&~~6.186(-2)&~~-2.777(-3)&\\
&$U^{238}$ &~~4.833(-1)&~~1.927(-1)&~~-1.283(-1)&~~-6.762(-3)&~~2.233(-3)&~~-1.536(-4)&\\
&$Pu^{239}$ &~~4.757 &~~-5.392&~~2.563&~~-6.596(-1)&~~7.820(-2)&~~-3.536(-3)&\\
&$Pu^{241}$ &~~2.990 &~~-2.882&~~1.278&~~-3.343(-1)&~~3.905(-2)&~~-1.754(-3)&\\
\hline
\end{tabular}
\label{table3}
\end{center}
\end{table}

\begin{figure}
\begin{center}
\includegraphics[scale=.6]{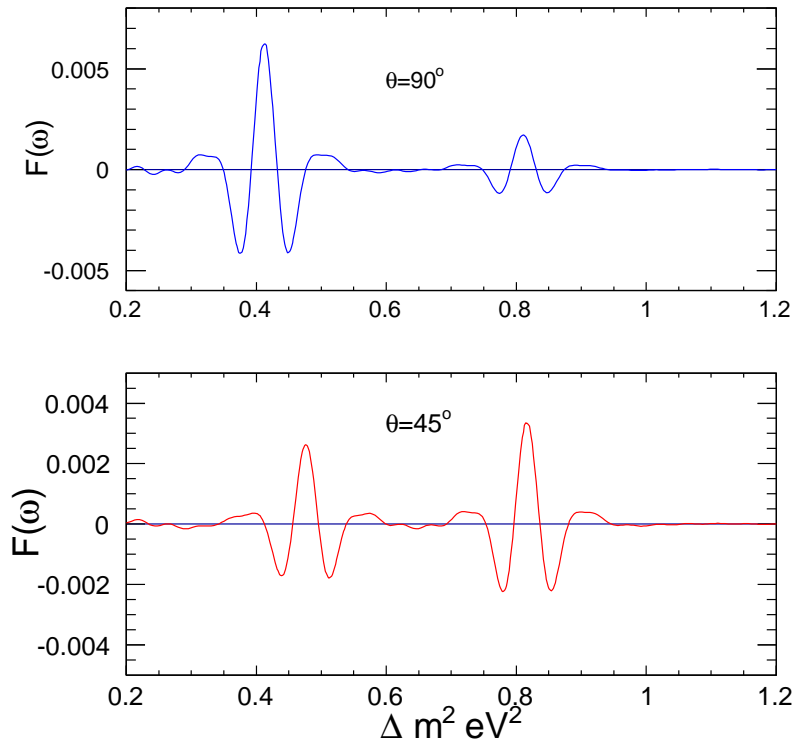}
\caption{The plot of the cosine Fourier transform (in arbitrary unit)
 as a function of $\Delta m^2$ at $\theta=\pi/2$ and $\theta=\pi/4$. The other parameters are as that of
fig \ref{fig_2d_3}. }
\label{fig_2d_4}
\end{center}
\end{figure}

The differential yield at energy $E$ and distance $L$ can
be written as,

\begin{eqnarray}
Y(L,E)= \Phi(E) \sigma(E) P_{e e} (L,E),
\end{eqnarray}

where $E$ is the energy of reactor anti-neutrino, $\sigma(E)$ is the interaction cross section
of anti-neutrino with matter and $P_{e,e}$ is the anti-neutrino survival probability as defined in 
Eq. \ref{p3}.
The leading order expression for the
cross section of inverse-$\beta$ decay ($\bar \nu_e \rightarrow  e^{+} + n$) is given by \cite{Vogel},

\begin{eqnarray}
\sigma = 0.0952 \times 10^{-42}~cm^2~(E_ep_e/~1MeV^2).
\end{eqnarray}

where $E_e = E_{\bar \nu} -(M_n-M_p)$ is the positron energy when neutron recoil
energy is ignored and $p_e$ is the positron momentum. 
The fractional contributions 
of $U^{235}:U^{238}:Pu^{239}:Pu^{241}$ to the total power are taken in the ratio $0.538:0.078:0.328:0.056$ respectively. We consider
two sterile mass states corresponding to the parameters
$m_1=0.052$ eV and $R=3.1 \times 10^{-7}$ m. This corresponds to 
$\Delta m_{14}^2 \sim 0.42$ $eV^2$, $\Delta_{15}^2 \sim 0.82$ $eV^2$ when $\theta=\pi/2$
and $\Delta m_{14}^2 \sim 0.48$ $eV^2$ , $\Delta_{15}^2 \sim 0.82$ $eV^2$ when $\theta=\pi/4$ (see table \ref{table1}).
In order to locate the mass peak, we consider the fourier cosine transform in the $1/E$ space given
by \cite{Zhan},

\begin{eqnarray}
F(\omega,L) = \int_{t_{min}}^{t_{max}} \big [Y(L,t)-Y_0(E,L) \big] cos(\omega t) dt,
\label{FT}
\end{eqnarray}

where $t=1/E$ which varies from $1/E_{max}$ to $1/E_{min}$ ($E_{max}=8$ MeV and $E_{min}=1.8$ MeV) and $\omega$ plays the role of frequency but in units of eV.
We define $Y_0(E,L)$ as the yield without $P_{ee}$ term in Eq. \ref{p3}. We have introduced $Y_0$ in Eq. \ref{FT} to improve
the sensitivity by substracting a background term. 
The fig. \ref{fig_2d_4} shows the cosine Fourier transform of the above spectrum as a function of $\Delta m^2=\omega/(2.54L)$
which shows sharp peaks when $\omega ~\sim 2.54 L \Delta m^2$. 
Since the active neutrino masses are nearly degenerate as compared to the sterile masses, the peaks appear at $\lambda_{mn}^2/R^2$.
When $\theta=\pi/2$, the two lowest modes are $(1,0)$ and $(1,1)$ corresponding to mass square difference of $0.42$ $eV^2$ and 
$0.82$ $eV^2$ respectively as shown in top panel. Although $F(\omega)$ is shown in arbitray units,
the height is proportional to the mixing probability. Since the height of first peak is more than the second, it would mean
$|U_{14}| > |U_{15}|$. The bottom panel shows the plot when $\theta=\pi/4$ corresponding to mass square 
differences of $0.48$ $eV^2$ and $0.82$ $eV^2$ respectively. In this case, the height of the second peak is more than the first one
resulting $|U_{15}| > |U_{14}|$. Although shown for $\theta=\pi/4$, it is noticed that in general
$|U_{15}| \ge |U_{14}|$ in the theta range $40^o < \theta < 60^0$ even though $\Delta m_{15}^2 > \Delta m_{14}^2$.

\begin{figure}
\begin{center}
\includegraphics[scale=.6]{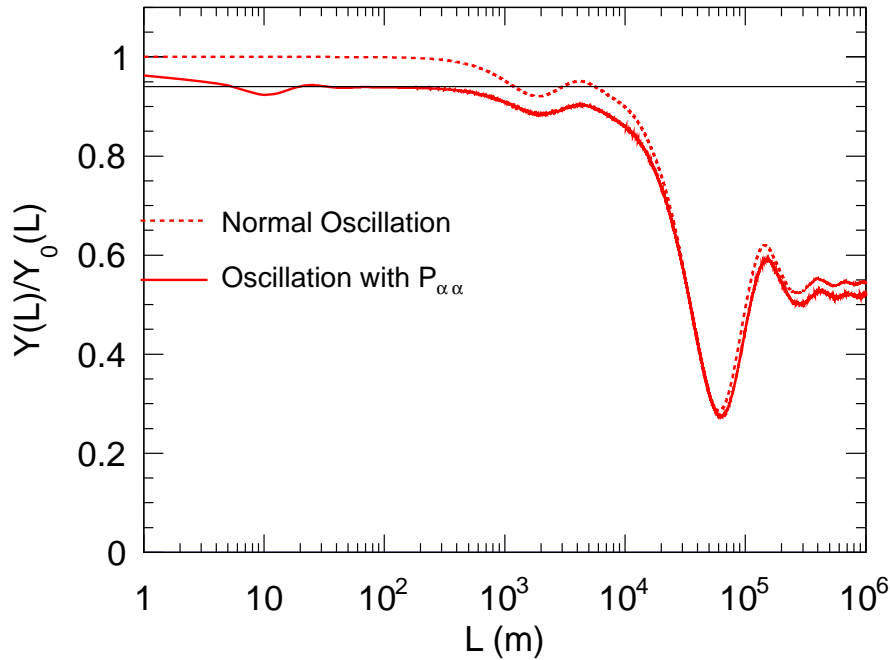}
\caption{The plot of the ratio of $Y(L)/Y_0(L)$ as a function of $L$. The dotted curve is obtained
using normal oscillation parameters i.e. the probability $P_{ee}$ is obtained without using $L^{0k}$
factor. The solid curve is obtained using $P_{ee}$ with $L^{0k}$ included.
This plot corresponds to $\theta = 45^{o}$, $R=3.1 \times 10^{-7}$, $m_1=0.052$ eV
with normal hierarchy. The line represents the average value at $0.94$. }

\label{fig_2d_5}
\end{center}
\end{figure}

Figure \ref{fig_2d_5} shows the ratio of the total yield $Y(L)/Y_0(L)$ as a function of $L$ in $m$.
The dotted curve is obtained using normal oscillation probability $P_{ee}$ which does not
include the active-sterile oscillation factor $L^{0k}$. 

The anti-neutrino survival probability is lowest when the argument in the exponential of
Eq. \ref{p3} is $\pi$. This corresponds to the relation,

\begin{eqnarray}
L \sim \frac{1.2 E}{\Delta m^2},
\label{G4}
\end{eqnarray}.

where we have replaced $\lambda^2$ by $\Delta m^2$. 
For normal oscillation, the dips occur at $L \sim 2000$ m and  
$L \sim 60000$ m corresponding to $\Delta m_{13}^2 = 2.42 \times 10^{-3}$ $eV^2$ and $\Delta m_{12}^2=7.45 \times 10^{-5}$ $eV^2$
respectively. 
This is consistent with the relation given by
Eq. \ref{G4} if we cnsider $<E> \sim 4$ MeV. When $L^{0k}$ is included, another dip occurs at $L \sim 10$ m corresponding to
$\Delta m^2=0.48$ $eV^2$. The effect due to other higher masses are not significant as the mixing probability decrerases with increasing
mass.
The black dotted line indicates the ratio at $0.94$ which is the average deficit reported in \cite{Muller}.

\section{Conclusions}
We have considered a toroidal extra dimensional space  associated with a shape moduli characterized by
an angle $\theta$ between the two large extra dimensions $R_1$ and $R_2$. The Kaluza-Klein compactification
results in a tower of bulk neutrinos which couple to the active neutrinos at the brane. 
The active-sterile mixing
probability depends strongly on the angle $\theta$ due to changing pattern of KK mas gaps resulting in 
level crossing. Considering only
the first two KK mass states corresponding to $(1,0)$ and $(1,1)$ modes in analogy with $(3+2)$ neutrino
mixing model, it is shown that there exists a range in $\theta$ $(\sim 40^o < \theta < \sim 60^o)$  where the mass of the 
higher $(1,1)$ KK mode is lower as compared to the mass of the $(1,0)$ or $(0,1)$ mode. Since the $(0,1)$ and
$(1,0)$ modes are degerate, it results in a higher active-sterile mixing probability for $(1,0)$ mode
as compared to the $(1,1)$ mode. In $(3+2)$ analogy, this would mean $|U|_{e5} > |U_{e4}|$ even though
$\Delta_{15}^2 > \Delta_{14}^2$. This is an important observation which can be verified from the 
short base line neutrino measurements, although present global anlysis seems to support the
above observation at $\theta \sim \pi/4$.
 The fourier analysis of the reactor anti-neutrino spectra at SBL also shows more qualitatively 
the above features which may also be possible to verify in near future with precision measurements.


\begin{thebibliography}{99}

\bibitem{Dienes_prl1}
K. R. Dienes, Phys. Rev. Lett. {\bf 88}, 011601 (2001).

\bibitem{Dienes_prl2}
K. R. Dienes and A. Mafi, Phys. Rev. Lett. {\bf 88}, 011602 (2002).


\bibitem{ADD1}
N. Arkani-Hamed, S. Dimopoulos and G. R. Dvali, Phys. Lett B {\bf 429}, 263  (1998).

\bibitem{ADD2}
N. Arkani-Hamed, S. Dimopoulos and G. R. Dvali, Phys. Rev. D {\bf 59}, 086003 (1999).


\bibitem {muller} Th. A. Muller et al, Phys. ReV. C {\bf 83}, 054615, (2011).

\bibitem {huber} P. Huber, Phys. ReV. C {\bf 84},  024617 (2011).

\bibitem{Kopp_prl}
J. Kopp, M. Maltoni and T. Schwetz, Phys. Rev. Lett. {\bf 107}, 091801 (2011).

\bibitem{Kopp1}
J. Kopp, P. A. N. Machado, M. Maltoni and T. Schwetz, J. High Energy Physics {\bf 05}, 0501 (2013).


\bibitem{Anton} 
I. Antoniadis, N. Akrani-Hamed, S. Dimopoulous and G. R. Dvali, Phys. Lett. B {\bf 436}, 257 (1998).

\bibitem{ADD3}
N. Arkani-Hamed, S. Dimopoulos, G. R. Dvali and J. March-Russell, Phys. Rev. D {\bf 65}, 024032 (2002).
\bibitem{Dienes}
K. R. Dienes, E. Dudas and T. Gherghetta, Nucl. Phys. B {\bf 557}, 25 (1999).
\bibitem{Dvali}
G. R. Dvali and A. Y. Smirnov, Nucl. Phys. B {\bf 563}, 63 (1999).
\bibitem{Bar}
R. Barbieri, P. Creminelli and A. Strumia, Nucl. Phys. B {\bf 585}, 28 (2000).

\bibitem{Davo}
H. Davoudiasl, P. Langakker and M. Perelstein, Phy. Rev. D {\bf 65}, 105015 (2002).

\bibitem{Gin}
D. M. Gingrich, Int. Journal of Moder Physics A {\bf 24}, 5173 (2009).
\bibitem{Cao}
Q. Cao, S. Gopalakrishna and C. P. Yuan, Phys. ReV. D {\bf 69}, 115003 (2004).
\bibitem{Gonzalez}
V. S. Basto-Gonzalez, A. Esmaili and O. L.  G. Peres, Phys. Lett B {\bf 718}, 1020 (2013).

\bibitem{Rode}
W. Rodejohann and H. Zang, Phys. Lett. B {\bf 737}, 81 (2014).

\bibitem{Dudas} E. Dudas, C. Grojean and S. K. Vempati, hep-ph/051100 (2005).

\bibitem{Garcia}
M. Gonzalez-Garcia, M. maltoni, J. Salvao, T. Schwetz, J. High Energy Physics {\bf 1212} 123 (2012).

\bibitem{Esmaili} A. Esmaili, O.  L. G. Peres and Z. Tabrizi, J. Cosmologyand astroparticle Physics 
            {\bf 12} 002 (2014).  

\bibitem{Vogel} P. Vogel and J. F. Beacom, Phy. Rev. D{\bf 60}, 053003 (1999), hep-ph/9903554.

\bibitem {Zhan} L. Zhan, Y. Wang, J. Cao and L. Wen, Phys. Rev. D{\bf 78}, 111103 (2008), hep-ex/0807.3203.

\bibitem{Giu}
E. Giusarma, M. Gerbino, O. Mena, S. Vagnozzi, S. Ho and K. Freese, astro-phi/1605.04320.



\end{thebibliography}
\end{document}